\documentstyle[12pt]{article}
\input{epsf}
\catcode`\@=11
\newcount\hour
\newcount\minute
\newtoks\amorpm
\hour=\time\divide\hour by60
\minute=\time{\multiply\hour by60 \global\advance\minute by-\hour}
\edef\standardtime{{\ifnum\hour<12 \global\amorpm={am}%
        \else\global\amorpm={pm}\advance\hour by-12 \fi
        \ifnum\hour=0 \hour=12 \fi
        \number\hour:\ifnum\minute<10 0\fi\number\minute\the\amorpm}}
\edef\militarytime{\number\hour:\ifnum\minute<10 0\fi\number\minute}
\def\draftlabel#1{{\@bsphack\if@filesw {\let\thepage\relax
   \xdef\@gtempa{\write\@auxout{\string
      \newlabel{#1}{{\@currentlabel}{\thepage}}}}}\@gtempa
   \if@nobreak \ifvmode\nobreak\fi\fi\fi\@esphack}
        \gdef\@eqnlabel{#1}}
\def\@eqnlabel{}
\def\@vacuum{}
\def\draftmarginnote#1{\marginpar{\raggedright\scriptsize\tt#1}}
\def\draft{\oddsidemargin -.2truein
        \def\@oddfoot{\sl preliminary draft \hfil
        \rm\thepage\hfil\sl\today\quad\militarytime}
        \let\@evenfoot\@oddfoot \overfullrule 3pt
        \let\label=\draftlabel
        \let\marginnote=\draftmarginnote
   \def\@eqnnum{(\theequation)\rlap{\kern\marginparsep\tt\@eqnlabel}%
\global\let\@eqnlabel\@vacuum}  }
\relax

\renewcommand{\theequation}{\arabic{section}.\arabic{equation}}

\def\be{\begin{equation}}
\def\ee{\end{equation}}
\def\bs{\begin{subequations}}
\def\es{\end{subequations}}

\renewcommand{\gg}{\rangle\!\rangle}
\renewcommand{\vert}{|\!|}

\begin{document}

\begin{titlepage}
\begin{flushright}
hep-th/0411067
\end{flushright}
\vspace{.7in}

\begin{center}
{\Large\bf Loop Operators and the Kondo Problem} \vskip .5cm
\vspace{.2in}

{\large Constantin  Bachas$\ ^{1,2}$ and Matthias R.\ Gaberdiel$\ ^2$}
\vspace{.2in}

{$\ ^1$ \em Laboratoire  de Physique Th{\'e}orique de l'Ecole
Normale Sup{\'e}rieure\footnote{\small Unit\'e Mixte du CNRS et
de l'Ecole Normale Sup\'erieure, UMR 8549}\\ 24 rue Lhomond, 75231
Paris cedex 05,
France\\ $\, $   \\
$\ ^2$  Institut f\"ur Theoretische Physik, ETH  H{\"o}nggerberg \\
8093 Z\"urich, Switzerland\\ $\, $   \\
{\rm \small Email: bachas@lpt.ens.fr ; gaberdiel@itp.phys.ethz.ch} 
}
\end{center}

\vskip .7in

\begin{center} {\bf ABSTRACT }
\end{center}
\begin{quotation}\noindent
We analyse the renormalisation group flow for D-branes in WZW models
from the point of view of the boundary states. To this end we
consider loop operators that perturb the boundary states away from
their ultraviolet fixed points, and show how to regularise and
renormalise them consistently with the global symmetries of the
problem. We pay particular attention to the chiral operators that
only depend on left-moving currents, and which are attractors of the
renormalisation group flow.  We check (to lowest non-trivial order
in the coupling constant) that at their stable infrared fixed points
these operators measure quantum monodromies, in agreement with
previous semiclassical studies.  Our results help clarify the
general relationship between boundary transfer matrices and defect
lines, which parallels the relation between (non-commutative) fields
on (a stack of) D-branes and their push-forwards to the target-space
bulk.
\end{quotation}
\vskip 1cm

\begin{flushleft}
LPTENS/04-46\\
November 2004\\
\end{flushleft}
\end{titlepage}
\vfill
\eject

\def\baselinestretch{1.2}
\baselineskip 14 pt

\noindent

\setcounter{section}{0}
\section{Introduction}

 The Kondo effect [1--6],
{\it i.e.} the screening of magnetic impurities
by the conduction electrons in a metal, 
has played a key role in
the early development of the renormalisation group. 
More recently
it has also proved instrumental in unraveling the elegant
structure of boundary conformal field theory. In modern language, the
screening describes the RG flow between the Cardy states
\cite{Cardy:1989ir} of the $\hat{su}(2)_k$ WZW  model, where $k$
is the number of electron channels that participate in the process
---  see \cite{Affleck:1995ge, Saleur:1998hq} for reviews 
and further references.

  In string theory  conformal boundary states correspond to D-branes, 
and this has revived interest in the problem from a different,
more geometrical point of view.\footnote{See, however,  also
\cite{Pradisi:1995qy} for an  earlier, algebraic investigation.} 
It has been appreciated, in
particular,  that the symmetric WZW  branes  wrap (twisted)
conjugacy classes of the group manifold
\cite{Alekseev:1998mc,Felder:1999ka}, that they are solutions of
the Dirac-Born-Infeld equations
\cite{Bachas:2000ik,Pawelczyk:2000ah,Bordalo:2001ec}, and that they 
are classified by an appropriate version of K-theory 
[16--20].\footnote{These references build upon the earlier 
work in \cite{Minasian:1997mm,Witten:1998cd,Bouwknegt:2000qt}. }
It was furthermore
argued that the RG flow between different (stacks of) D-branes
in these models is a process
that lowers the energy by switching on certain non-abelian
worldvolume backgrounds \cite{Alekseev:2000fd,Alekseev:2002rj}. This
is similar to the dielectric effect for  D-branes in the presence of a
non-vanishing Ramond-Ramond field strength \cite{Myers:1999ps}. 

   In a separate development  \cite{Felder:1999cv}  it was suggested
that the choice of  conformal boundary conditions  in the WZW
model  amounts to the insertion of (spacelike)  Wilson loops   in
the three-dimensional Chern-Simons theory. The  approach was then
generalised \cite{Fuchs:2002cm}  to all  rational CFTs, with
Wilson loops  replaced by  more general `topological defect
lines'. These were introduced in ref.\ \cite{Petkova:2000ip} as
formal bulk operators that commute with the chiral algebras of the
theory. They are special cases of conformal defects, which in
general only commute with the diagonal Virasoro algebra
\cite{Oshikawa:1996dj,Bachas:2001vj,Quella:2002ct,Saleur:1998hq}.
Topological defects have been recently  used  to conjecture relations
between boundary RG flows  \cite{Graham:2003nc}, and  as generators of 
generalised Kramers-Wannier duality transformations
\cite{Frohlich:2004ef}.

  Our aim in the present work will be to  clarify
the relation between these different approaches to the Kondo
problem. We will work in the boundary state formalism
\cite{Callan:1987px,Polchinski:1987tu}, and consider the
renormalised transfer-matrix operator that perturbs the theory
away from its ultraviolet fixed point. We will argue  that  this
operator can be pushed forward smoothly into the bulk, so that it 
depends on left-moving currents only. As a result, it commutes
with the right-moving  algebra  along the entire RG flow. At the
infrared fixed point, this bulk  operator also  commutes  with  the
left-moving algebra, and can  be identified with the topological 
Wilson loops of ref.\ \cite{Felder:1999cv}, or with the
topological defect lines of ref.\ \cite{Petkova:2000ip}. The
fixed-point operator is in fact the trace of the quantum monodromy
matrix [37--40].
Though we do not have a full proof, we will verify these
statements explicitly at low orders in the perturbative expansion.

  The existence of renormalisable loop operators, which can be moved
freely to the boundary from the bulk, is a remarkable feature of
the WZW model. It implies universal RG flows, common to all
ultraviolet fixed points. From the geometric point of view, it
means that the corresponding D-brane fields admit special,
universal push-forwards to the entire target manifold. It would
be very interesting to understand whether and, if so, how these
features could be generalised to other models.

  Besides those cited above, several other papers touch upon various
aspects of our work. Most closely related is ref.\
\cite{Hikida:2001py}, whose authors studied the  
same transfer-matrix operator  and  its action on  WZW boundary
states.  Their analysis is, however, semiclassical, and they do
not try to define the operator in the bulk, nor away from its
critical point. The  connection between quantum monodromy matrices
and boundary flows has also been discussed by Bazhanov {\it et.al.}\ 
\cite{Bazhanov:1994ft} in the context of minimal models. Their
conformal perturbation scheme could be used in the Kondo problem,
though it would force one to abandon the explicit $su(2)$
symmetry of the flow.\footnote{We thank  Sasha Zamolodchikov for a
discussion of this point.} Finally, other discussions of boundary
states for D-branes with non-abelian backgrounds switched on can
be found in refs.\ \cite{Ishibashi:1998ni} and \cite{Pesando:2003ww}.  

The plan of this paper is as follows~: in section 2 we discuss
the push-forward of D-brane fields to the bulk of the target space,
and the corresponding push-forward of the boundary transfer
matrix to a loop operator in the worldsheet bulk. We also 
explain the difference between conformal, chiral and topological
defects. In section 3 we introduce the symmetric defect lines of the
WZW model, and give a semiclassical argument that identifies their 
fixed points. In section 4 we regularise these defects,  so as to
respect a number of classical symmetries, and  expand the
regularised loop operator up to fourth order in the coupling
constant. In section 5 we renormalise this operator and study its
infrared fixed points. We verify explicitly at low orders
that at the stable infrared fixed points, the  operator commutes
with the current algebra, and that its spectrum is given 
by generalised quantum dimensions. Finally in section 6 we pull back 
these operators to the boundary, and discuss the induced boundary RG
flow. We end with some open questions.

\setcounter{equation}{0}
\section{Transfer matrices and defect lines}

 Consider a stack of $n$  identical  D-branes in a target space
parametrised by  $X^M$. Let  $n \vert {\cal B}\gg\ $ be the conformal
boundary state of the branes, with  $  A$ and  $  Y$  their
matrix-valued gauge and  coordinate fields. Switching on non-zero
backgrounds for these fields leads to the following (formal)
modification of the boundary state~:
\begin{equation}\label{defn}
n\;\vert {\cal B}\gg\ \longrightarrow \;
{\rm Tr} \;  {\rm P} \; 
{\rm exp}\left( i \int_0^{2\pi}\hskip -0.1mm d\sigma \left[ A_M(X)\;
\partial_\sigma X^M + {1\over 2\pi
\alpha^\prime}\;   Y_M(X)\;
\partial_\tau X^M\right]  \right) \; \vert {\cal B}\gg\ .
\end{equation}
Here the integral runs over the boundary of the half-infinite
cylinder that is parametrised by $\sigma\in [0,2\pi]$ and 
$\tau\in [0,\infty)$, the trace is over the Chan-Patton indices of 
$A$ and $Y$, and P denotes path ordering. The factor 
${2\pi \alpha^\prime}$, conventional in the string-theory literature, 
could have been absorbed in the definition of $Y$. The
worldsheet fields $X^M(\sigma,\tau)$ are 
quantised in the closed-string channel, {\it i.e.}\  with $\tau$ being
the time-like coordinate. The boundary at $\tau=0$ is thus spacelike
and (barring singularities at coincident points) all worldsheet
fields in expression (\ref{defn}) commute. The path ordering is, of
course, still non-trivial because the background fields are
matrix-valued. 

  The attentive reader will have noticed that we wrote
$A$ and $Y$ as one-form fields, defined on the entire target
manifold, ${\cal M}$, even though they live, a priori, only in the 
tangent (or normal) bundle of the D-brane worldvolume. What
we implicitly assume is that the freedom in choosing this 
`push forward' does not matter, because of the boundary
conditions imposed by $\vert {\cal B}\gg$.  Consider for example
a D-instanton at the origin, whose boundary state satisfies the
conditions  $X^M(\sigma, 0) \vert {\cal B}\gg\, =\, 0$. Ignoring
problems of operator ordering, one would conclude that
the integrand in (\ref{defn})  reduces to 
$ Y_M(0)  \partial_\tau   X^M/2\pi \alpha^\prime $, so that only the
value of $Y$ at the origin really matters. This is of course a
semiclassical argument, and it is possible that it fails at the
quantum level. 

  A closely-related question is the following. The path-ordered
exponential in eq.\ (\ref{defn}) is the closed-string dual of the
transfer matrix, or evolution operator in the open channel. Let us
denote this operator by ${\cal O}(A,Y)$. If we were to quantise $X^M$
with $\sigma$ as time, then ${\cal O}$ would be the evolution operator
in the interaction representation, around the theory defined by the
boundary condition ${\cal B}$.\footnote{This was worked out
explicitly for plane-wave backgrounds in  \cite{Bachas:2003sj}, using
results of \cite{Bachas:2002jg,Hikida:2003bq}. The fact that 
the interaction Lagrangian may depend on the time derivatives of
fields is a subtle complication  that does not invalidate our
assertion \cite{Matthews}.} From this perspective, only the action 
of ${\cal O}$ on the  boundary state $\vert {\cal B}\gg$ is really
specified. Extending the action of the path-ordered exponential to
{\it all} closed-string states requires a push-forward of $A$
and $Y$ to the whole target manifold. This makes it, in turn, 
possible to push the integration contour from the boundary to the
interior of the worldsheet, where boundary conditions are no more
effective (see figure 1). Reversing this operation amounts to taking
the limit
\begin{equation}\label{limit}
{\rm lim}_{\varepsilon\to 0}\; {\cal O}(A,Y)\; 
e^{i\varepsilon (L_0+\bar L_0)} \vert  {\cal B}\gg\ ,
\end{equation}
where $L_0+\bar L_0$ is the 
Hamiltonian in the closed-string channel. If the limit is
non-singular, it should only depend on the pullbacks of $A$ and $Y$ 
to the tangent and normal bundles of the original, unperturbed D-brane.

\vskip 0.3cm

\begin{figure}[ht]
       \hbox{\epsfxsize=90mm%
       \hfill~\hskip 2.5cm 
       \epsfbox{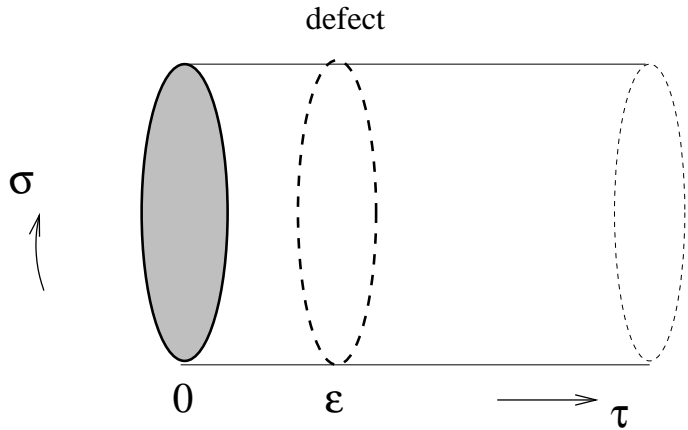} \hfill~}
       \caption{\small A push-forward of the D-brane fields to the
       target manifold,  makes it possible to
       push  the integration contour of  (\ref{defn}) from the
       boundary to the interior of the worldsheet.}
\end{figure}

\vskip 0.3cm

  In the open channel, the operator ${\cal O}(A,Y)$ describes the
interactions of a point defect sitting in the interior of the
worldsheet. Such defects have been previously discussed in the
condensed-matter \cite{Oshikawa:1996dj,Saleur:1998hq} and in the
string-theory \cite{Bachas:2001vj,Quella:2002ct} literature. 
By folding the worldsheet along their worldline, one may picture bulk
defects geometrically as  D-branes in the tensor product,  
${\cal M}\times {\cal M}$, of two identical target manifolds. 
The trivial defect, {\it i.e.}\ the identity operator, corresponds 
to the  diagonal brane in ${\cal M}\times {\cal M}$
\cite{Bachas:2001vj,Quella:2002ct}. Its geometric and gauge degrees of
freedom are precisely given by two one-form fields living  on 
${\cal M}$. This fits very nicely with the fact that $A_M(x)$ and
$Y_M(x)$ account (together with the tachyon field) for  the most
general renormalisable couplings on a defect worldline.

  Conformal defects are obtained at the fixed points $(A^*,Y^*)$ of
the RG flow. The corresponding loop operators commute with the
generators of conformal transformations that preserve the defect
worldline, 
\begin{equation}\label{cnds}
\left[\; {\cal L}_n  \; ,  \; {\cal O}(A^*,Y^*)\; \right]\; =  0\ \
\ \ \ \forall \; n \ ,
\end{equation}
where ${\cal L}_n = L_n - \bar L_{-n}\ $  with $L_n$ and 
$\bar L_n$  the left- and right-moving Virasoro generators,
respectively. [Note in particular that ${\cal L}_0$ generates
translations in $\sigma$.] It follows from conditions (\ref{cnds})
that if $\vert{\cal B}\gg$ is a conformal boundary state, {\it i.e.}\
if it is annihilated by all ${\cal L}_n$, then 
${\cal O}(A^*,Y^*) \vert{\cal B}\gg$ is also conformal. Thus, if the
limit (\ref{limit}) were  smooth, pulling back the RG flow of a bulk
defect  to {\it any}  conformal boundary would induce a (possibly
trivial)  RG flow of  the boundary state. Conformal defect lines, in
particular, act as solution-generating symmetries, {\it i.e.}\  they
map conformal to conformal  boundary states.

  In general, taking the limit (\ref{limit}) can  be tricky, since
the presence of the  boundary could conceivably modify the RG flow
at distances much greater than $\varepsilon$. There exists, however,
one special class of defects that commute with the 
right-moving Virasoro algebra,
\begin{equation}
\left[\; {\bar L}_n  \; ,  \; {\cal O}_{\rm chir}(A,Y)\; \right]\; =
0\ ,
\end{equation}
and for which the limit (\ref{limit}) is trivial. We will refer to
them  as chiral defects (or anti-chiral if they commute with the
left-moving algebra). Chiral defects need not be conformal, but
they can be regularised and renormalised in a way that preserves
their invariance under $\sigma$-translations. They therefore
commute with both ${\cal L}_0$ and $\bar L_0$, and hence also with
the closed-string Hamiltonian. Thus, chiral defects  can be taken
freely to the boundary of the cylinder, where they induce RG
flows of boundary states.

  Defect lines that are both conformal and chiral correspond to
operators that commute with both the left  and the right Virasoro
algebras, 
\begin{equation}\label{topo}
\left[\; {L}_n  \; ,  \; {\cal O}_{\rm chir}(A^*, Y^*)\; \right]\; =
\left[\; {\bar L}_n  \; ,  \; {\cal O}_{\rm chir}(A^*,Y^*)\;
\right]\; = 0\ .
\end{equation}
We will call such defect lines topological, since their action only
depends on the homotopy class of the integration contour. The
corresponding operator will be denoted  for short by
${\cal O}_{\rm top}$. The conditions (\ref{topo}),  together with a
Cardy constraint of integral multiplicities, have been used in
ref.\ \cite{Petkova:2000ip} to give an algebraic definition of
(topological)  defect lines. Notice that the Cardy constraint is
automatically obeyed provided ${\cal O}_{\rm top}$ is the transfer
matrix for some  local defect. Note also that ${\cal O}_{\rm top}$
need not be continuously connected (in D-brane configuration space) to
the identity operator. What can be asserted, following
ref.\ \cite{Graham:2003nc}, is that if 
$\vert {\cal B}\gg \to \vert {\cal B}^\prime\gg$ is an allowed
boundary RG flow, then so is 
${\cal O}_{\rm top} \vert {\cal B}\gg \to {\cal O}_{\rm top} \vert
{\cal B}^\prime\gg$  for any topological-defect operator. This
observation gives support for the flows that have been proposed in 
\cite{FS1,Fredenhagen:2003xf}.

  Simple examples of topological defect operators are the generators
of continuous symmetries, 
${\cal O}_{\rm top}  = {\rm exp}(i\lambda\oint J)$ with 
$\lambda$ a real  parameter and $J$ any abelian chiral spin-$1$
current. Another set includes discrete  automorphisms of the CFT, as
analyzed in ref.\ \cite{Frohlich:2004ef}. Both of these types of
operators are invertible, and do not arise as infrared fixed points of
RG flows, in contrast to the Kondo problem operators that we will
study here.


\setcounter{equation}{0}
\section{Semiclassical analysis of WZW defects}

  One may identify  conformal defects in the WZW model by a
semiclassical argument,  analogous  to the one 
used to argue for the existence of a conformal theory in the bulk
\cite{Witten:ar}. We recall the expressions for the  left- and
right-moving  currents of the model,
\begin{equation}
J(x^+)  = -i \kappa\; (\partial_+  g)  g^{-1} \ \ \ {\rm and} \ \ \ \
\bar J(x^-) =   i \kappa\;  g^{-1}\partial_- g \ ,
\end{equation}
where $x^\pm = \tau\pm\sigma$, and $g(x^+,x^-)$ takes values in  some
(simple compact) Lie group $G$.  The parameter $\kappa$ is related to
the integer level $k$ through 
\begin{equation}
\kappa = \psi^2 k/2\ ,
\end{equation}
with $\psi$ the length of long roots of the Lie algebra $g$ of $G$.
The currents generate the independent left and right symmetry
transformations
\begin{equation}
g \to u(x^+)^{-1}\; g\; \bar u(x^-)\ ,
\end{equation}
 under which they themselves transform  as~:
\begin{equation}\label{trans}
J \to u^{-1} J u + i\kappa\; u^{-1} \partial_+  u \qquad
\ {\rm and} \qquad
\bar J \to \bar u^{-1}  \bar J  \bar u + i\kappa\;
\bar u^{-1} \partial_-  \bar u\ .
\end{equation}
The Poisson bracket algebra of the Fourier moments, $J_n^a$, of
these currents is the classical counterpart of the standard left-
and right-moving affine Kac-Moody algebras $\hat{g}$ with central
extension $k$. This algebra implies the transformation rules
(\ref{trans}), and vice versa.

  We will be interested in bulk defects which interact linearly
with the currents of the model, and preserve a global $G$ symmetry. The
classical observable that corresponds to the quantum transfer matrix
of such  defects reads\footnote{There exist in fact more general
defects that respect a global $G$ symmetry, but their classical
observables are sums of products of the  above basic ones.} 
 \begin{equation}\label{basic}
{\cal O}_{\omega}(\lambda,\bar\lambda; R ) =  {\rm Tr}_R\; {\rm P \; exp}
\;\left( i  \int_0^{2\pi}  d\sigma\;  ( \lambda\; J^a   \; - \;
\bar \lambda\;  \omega(\bar J^a) ) \; t^a \right)\ .
\end{equation}
Here $t^a$ are the generators of $g$ in an $n$-dimensional
irreducible representation $R$, $\omega$ is an automorphism of the
Lie algebra $g$, $\lambda$ and $\bar\lambda$ are independent real
coupling constants, and there is an implicit sum over the adjoint
index $a$.  It will be useful to consider (\ref{basic}) as the
Wilson loop of a two-dimensional gauge field with components
\begin{equation}
(\alpha_+,\alpha_-)\;=\; (\; \lambda\; J^a t^a\; ,\;
\bar\lambda\; \omega(\bar J^a)t^a\; )\  .
\end{equation}
By virtue of the non-abelian Stokes theorem, the observables
(\ref{basic}) are conserved under the  (closed-string) time
evolution, provided  that the field strength 
$F_{+-}= i\;[\alpha_+,\alpha_-]$ vanishes. This is however  not the
case in general, except for chiral (or antichiral) defects for which
the coupling $\bar\lambda$ (or $\lambda$) is zero.

Let us concentrate on the chiral defects, with Wilson loops
\begin{equation}\label{bas}
{\cal O}_{\rm chir}(\lambda ; R ) =  {\rm Tr}_R\; {\rm P \; exp}
\;\left( i  \lambda \oint_C dx^+ \; J^a t^a \right)\ .
\end{equation}
Classically, such Wilson loops are topological, {\it i.e.} they only
depend on the homotopy class of the contour $C$. This property does
not, a priori, survive the renormalisation process, which introduces 
through dimensional transmutation a length scale. There is however
one special value of the coupling,
\begin{equation}
\lambda^* = -{1\over \kappa}\ ,
\end{equation}
for which ${\cal O}_{\rm chir}(\lambda^*; R)$ is in fact invariant
under the full symmetries (\ref{trans}) of the WZW model. 
To see why, recall that a Wilson loop is invariant under the gauge
transformations 
\begin{equation}
\alpha_\pm \to v^{-1}\alpha_\pm v - i v^{-1}\partial_\pm v
\end{equation}
for  arbitrary $v(x^+,x^-)$. The transformation (\ref{trans}) of
$\ (\lambda^* J^a t^a \;,\; 0)\ $ is of precisely this same form
with $v(x^+,x^-)=u(x^+)$. An immediate consequence is that this
special Wilson loop has vanishing Poisson brackets with both current
algebras, 
\begin{equation}
\{ J_n^a\;,\; {\cal O}_{\rm chir}(\lambda^* ; R ) \}_{\;_{\rm PB}}
= \{ \bar J_n^a\;,\; 
{\cal O}_{\rm chir}(\lambda^* ; R )  \}_{\;_{\rm PB}} = 0\ .
\end{equation}
As we will verify in the following sections, to lowest order in
$1/\kappa$ these relations do survive the canonical correspondence 
$\{\; , \; \}_{\;_{\rm PB}} \to[\; , \; ]\ $ (though in general
$\lambda^*$ has a finite, scheme-dependent renormalisation).
Since the Virasoro generators are quadratic in the affine
currents, ${\cal O}_{\rm chir}(\lambda^* ; R )$ will thus also obey 
eq.\ (\ref{topo}) that characterises topological defects.

   The above semiclassical argument can help us identify another class 
of defects, which are conformal but not topological. They correspond
to the symmetric choice
\begin{equation}
\lambda = \bar\lambda = {\lambda^*\over 2}\ .
\end{equation}
With this choice, the observables (\ref{basic}) are invariant under 
the (vector-like) transformations that have 
$u(x) = \omega(\bar u (-x))$, as the reader can easily verify.
These vector-like transformations are generated by the linear
combinations
of  currents
\begin{equation}
{\cal J}_n^a = J_n^{a} +  \omega (\bar J_{-n}^a)  \ ,
\end{equation}
which must therefore have vanishing Poisson brackets with the loop
observables,  
\begin{equation}\label{dig}
\left\{ {\cal J}_n^a \;,\; 
{\cal O}_\omega \left({\lambda^*\over 2}, {\lambda^*\over 2} ; R
\right) \right\}_{\;_{\rm PB}} = 0\ \ .
\end{equation}
Note that the ${\cal J}_n^a$ define a current algebra without central 
extension. Now assuming that (\ref{dig}) survives in the quantum
theory, and recalling that 
the Virasoro generators are qua\-dratic in the currents and
$\omega$-invariant, we immediately conclude that 
${\cal O}_\omega (\lambda^*/2, \lambda^*/2 ; R )$ will obey the defining 
relations (\ref{cnds}) of conformal defects. Note, however, that
these defect lines are not topological --- they do not, in
particular, commute with the $\tau$-evolution. In fact, they
correspond, as we  shall later see, to unstable fixed points of the RG
flow. 

   The topological observables ${\cal O}_{\rm chir}(\lambda^* ;R)$,
together with their right-moving counterparts, measure the invariant
monodromies of classical solutions. More explicitly, a general
solution of the classical WZW equations factorises into the product of
a left- and a right-moving part, 
\begin{equation}\label{clal}
g(x^+,x^-) = g_+(x^+)^{-1}  g_-(x^-)\ .
\end{equation}
Since only $g$ needs to be a single-valued function on the worldsheet,
one has 
\begin{equation}
g_\pm(x^\pm \pm 2\pi) = M g_\pm(x^\pm) \ ,
\end{equation}
with $M$ an arbitrary but constant group element.\footnote{In WZW
orbifolds, left and right monodromies need only be equal up to
orbifold identifications.} As the reader can easily check,
\begin{equation}
{\cal O}_{\rm chir}(\lambda^*; R) = {\rm Tr}_R\;  M  \ ,
\end{equation}
where the trace is evaluated in the $R$ representation. Note that
the decomposition (\ref{clal}) is not unique --- the  freedom to
redefine $g_\pm =  g_0\;  g_\pm\ $, with $g_0$ an arbitrary
constant group element,  changes $M$  to $g_0 Mg_0^{-1}$.  This is
compatible with the fact that our topological observables only
measure the conjugacy class of the monodromy.

  Using the above freedom, one can bring $g_\pm$ to the canonical
form 
\begin{equation}
g_\pm =   e^{\pm i\Lambda\; x^\pm}\; \tilde g_\pm(x^\pm)\,,
\end{equation}
where $\tilde g_\pm$ are single-valued functions on the circle, 
$M = e^{2\pi i\Lambda}$ is the monodromy matrix, and $\Lambda$ can be 
chosen (by a Weyl reflection) to lie in a positive Cartan alcove
of $\hat{g}$. The triplet $(\tilde g_+, \tilde g_-, \Lambda)$ 
describes the classical phase space of the WZW model. A geometric
quantization of this phase space\footnote{The canonical quantisation 
of the corresponding boundary theory was recently studied in
\cite{Gawedzki:2001rm}.}, known as co-adjoint orbit or 
Kirillov-Konstant quantization
\cite{Alekseev:1990vr,Gawedzki:1990jc,Chu:1991pn} 
(see also \cite{Pressley:1988qk}), gives the following spectrum for
the monodromy~: 
\begin{equation}\label{hvee}
\Lambda  =  {\mu + \rho \over k+ h^\vee} \, \ \ \ \ {\rm in\  the\
sector}\ \ {\cal H}_\mu\otimes\bar{\cal H}_\mu\ .
\end{equation}
Here ${\cal H}_\mu$ and $\bar{\cal H}_\mu$ are integrable
highest-weight representations of the left and right current algebras,
respectively; they are labelled by a highest weight vector $\mu$ of
the corresponding representation of $g$. Furthermore, $\rho$ is the
Weyl vector of $g$, {\it i.e.}\ one half of the sum of all positive
roots of $g$. Semiclassical reasoning does not actually determine the
finite shift of $k$ by the dual Coxeter 
number $h^\vee$ --- a detailed quantum calculation is needed for
this. Using (\ref{hvee})  one finds the following spectrum for the 
topological defect operator,
\begin{equation}\label{topW}
{\cal O}_{\rm chir}(\lambda^\ast; R)   = {\rm Tr}_R \left( 
e^{2\pi i \Lambda} \right) = {S_{R\mu} \over S_{0\mu}} \, \ \ \ \ \ 
{\rm in\ \  }  {\cal H}_\mu\otimes\bar{\cal H}_\mu\ .
\end{equation}
Here $S$ is the modular transformation matrix of the chiral
characters, and $0$ (and $R$) denotes the Kac-Moody
representation built on the trivial (or the $R$) representation of
$g$. Note that the trace of $M$ in the above expression can be
computed for an arbitrary representation $R$, not only for those
corresponding to integrable representations of the current
algebra.\footnote{Representations that fall outside the Cartan alcove
of $\hat{g}$ can be related to integrable highest weight
representations by an affine Weyl transformation. The corresponding
traces are then equal (up to signs).}

    We end  this section with some remarks. Firstly, the topological
defect  lines (\ref{topW}) should `lift'  to the (spacelike) Wilson
lines of the 3d  Chern-Simons theory \cite{Felder:1999cv}, but it is
not clear whether a lift exists for the more general, non-topological
defects. Second, the above loop observables define a classical
(fusion) algebra, 
\begin{equation}
{\cal O}_\omega(\lambda, \bar\lambda  ; R_1 )\;  
{\cal O}_\omega(\lambda, \bar\lambda  ; R_2 )
= \sum_{R\in R_1\otimes R_2}  {\cal O}_\omega(\lambda, \bar\lambda  ; R )\ ,
\end{equation}
for any values of the coupling constants. It would be interesting to
know whether this algebra structure survives at the quantum level.
Finally, the symmetric  defects discussed here form a special class,
and do not exhaust all conformal defects of WZW models.


\setcounter{footnote}{1}
\setcounter{equation}{0}
\section{Regularised loop operators}

  As a first step towards quantising the defect lines of the
previous section, we  will now describe a regularisation scheme and
write ${\cal O}_\omega^{\rm reg}(\lambda,\bar\lambda; R)$ as an operator in
the enveloping algebra of the current algebras. We start by defining
the regularised left-moving currents 
\begin{equation}\label{jreg}
J_{\rm reg}^a(\sigma) = \sum_{n\in {\bf Z}} J^a_n\;  
e^{-in\sigma - |n| s/2} \,,
\end{equation}
where $s$ is a short-distance cutoff, and the $J^n_a$ are
generators of the  Kac-Moody algebra\footnote{Our conventions on
affine algebras follow those of  ref.\ \cite{GO}.}
\begin{equation}
[J^a_n, J^b_m] = i f^{abc}\;  J^c_{n+m} + {\kappa}\, n\; \delta^{ab}
\delta_{n+m,0}\ .
\end{equation}
A similar expression defines the regularised right-moving
currents. Note that this regulator preserves the classical algebra
of analytic symmetry transformations, {\it i.e.}\ transformations 
generated by positive-frequency modes of the currents.

   Let us concentrate first on chiral defects, and expand the loop
observables (\ref{bas}) in  a power series of  the coupling constant,
\begin{equation}\label{path}
{\cal O}_{\rm chir}(\lambda; R) =  \sum_{N=0}^\infty
{(i\lambda)^N }
\; {\cal O}^{(N)}(R) \ ,
\end{equation}
where
\begin{equation}\label{path1}
{\cal O}^{(N)}(R) =
{\rm Tr}_R\; (t^{a_1}\cdots t^{a_N}) \left( \prod_{i=1}^{N}
\int_0^{2\pi} 
d\sigma_i \right) \;\theta_{\sigma_1>\cdots >\sigma_N}\;
J^{a_1}(\sigma_1)\cdots J^{a_N}(\sigma_N)\;
\end{equation}
with   $\theta_{\sigma_1>\cdots >\sigma_N}=1 $  if
$\sigma_1>\sigma_2 \cdots >\sigma_N$, and 
$\theta_{\sigma_1>\cdots >\sigma_N}=0$ otherwise.
Classically,  the  order of the currents in the above expression is
irrelevant, but in the quantum theory a precise order must be
specified. To guide our  choice we will insist that the following two 
symmetries of  the classical observables be preserved:  (i) the path
can start at any  point  $\sigma_0$ on the circle, and (ii) the result
is invariant if we reverse the orientation of the loop,  and  change
$R$  with its  conjugate representation.\footnote{This is more 
familiar in  Yang-Mills theory, where  a  quark line
running  forward in time cannot be distinguished from an
antiquark line that goes  backwards. To check this symmetry remember
that if $t^a$ are the hermitean generators in the representation $R$,
then the generators of the conjugate representation are given by $(-t^a)^*$.}  
Neither of these symmetries would  be preserved if we kept the order
of the currents as in (\ref{path1}). However, these symmetries are
preserved if we average over the $N$ cyclic permutations, as
well as the $N$ permutations that are obtained from them by reversing
the order, {\it i.e.}\ by combining them with the  
permutation $\rho: j \mapsto N+1-j$. We thus define
\begin{eqnarray}\label{path2}
{\cal O}_{\rm reg}^{(N)}(R) =
{\rm Tr}_R\; (t^{a_1}&&
\hskip -0.5cm \cdots t^{a_N}) \left( \prod_{i=1}^{N} \int_0^{2\pi} 
d\sigma_i \right) \;\theta_{\sigma_1>\cdots >\sigma_N}\times\\ 
\nonumber
&& \qquad \qquad
\times {1\over 2N}  \left( J_{\rm reg}^{a_1}(\sigma_1)\cdots 
J_{\rm reg}^{a_N}(\sigma_N)\;
+ {\rm cyclic}\ + {\rm reversal} \right)\ .
\end{eqnarray}
Note that since the bare currents at non-coincident points
commute, the  choice of ordering is part of the regularisation
prescription for the loop operator. Note also that our prescription 
(which is not unique) guarantees that $ {\cal O}_{\rm reg}^{(N)}(R)$
commutes  with the generator  of  $\sigma$-translations. As explained
in the previous section, such chiral operators can  be transported
freely to the  boundary of the half-infinite cylinder.

 Plugging the mode decomposition (\ref{jreg}) in (\ref{path2}), and 
performing explicitly the integrals leads to the following
expressions for the first few values of $N$~:
\begin{equation}\label{R2}
{\cal O}_{\rm reg}^{(2)}(R)  = 2\pi^2\;
{\rm Tr}_R(t^a t^b)\;  J_0^a J_0^b\ ,
\end{equation}
\begin{equation}\label{R3}
{\cal O}_{\rm reg}^{(3)}(R) =   {2\pi^2\over 3} \; {\rm Tr}_R(t^a
t^b t^c) \left[ \, {\pi\over 3}\;   J_0^a  J_0^b J_0^c +
\sum_{n\not= 0 } {i\over  n} \tilde J^a_{-n} \tilde J^b_{n}
J^c_0\;  + {\rm cyclic} + {\rm reversal}\; \right]\ ,
\end{equation}
 and
\begin{eqnarray}
{\cal O}_{\rm reg}^{(4)}(R)&=& {\pi^2\over 2} \; {\rm Tr}_R(t^a
t^b t^c t^d)\; \Biggl[\, {\pi^2\over 6}\;   J_0^a  J_0^b  J_0^c
J^d_0 \,+ \,\sum_{n\not= 0} {i\pi \over  n} 
\tilde J^a_{-n} \tilde J^b_{n} J^c_0  J^d_0 
\nonumber \\
&& \quad + \sum_{n\not= 0 } {1 \over  n^2}\, \left( \tilde
J^a_{-n} \tilde J^b_{n} J^c_0  J^d_0 - \tilde J^a_{-n}
 J^b_{0}\tilde J^c_n  J^d_0\right)
\label{R4}  + \sum_{\stackrel{\footnotesize{m,l,n\not=
0}}{m+n+l=0}} {1\over ml}\, \tilde J^a_{m} \tilde J^b_{n}\tilde
J^c_l  J^d_0 \\
&& \quad - {1\over 2} \sum_{m,n\not= 0}{1\over mn} \tilde J^a_{-n}
\tilde J^b_{n}\tilde J^c_{-m} \tilde J^d_m
\;   +\;
{\rm cyclic} + {\rm reversal} \;   \Biggr]\ . \nonumber
\end{eqnarray}
Here we have used the short-hand notation 
$\tilde J^a_n = J^a_n \;e^{- |n| s/2}$. Note that in  
deriving these formulae it helps to sum over all cyclic permutations
of the currents and use the cyclic property of the trace, before
performing  explicitly the $\sigma_j$ integrals. 

 Let us pause for a moment to describe our group theoretic
conventions. The $t^a$ are (traceless) generators of $g$ in the
representation $R$, which we assume irreducible. The Killing form is
$\delta^{ab}$, so adjoint indices can be raised and lowered freely. 
We denote by ${\rm dim}(R)$ the dimension of $R$, and by 
$C(R)$ the value of the quadratic Casimir operator in the
representation $R$, {\it i.e.}\ 
$\sum_a t^a t^a = C(R) \times{\rm identity}$. In these conventions   
\begin{equation}
{\rm Tr}_R\;(t^a t^b) = I_R \, \delta^{ab}\ , \ \ \   {\rm where}\
\ \ \ I_R = {C(R)\times  {\rm dim}(R)  \over {\rm dim}(g)}\ ,
\end{equation}
where ${\rm dim}(g)$ is the dimension of the adjoint representation.
Furthermore, the dual Coxeter number
$h^\vee$ is given by~:
\begin{equation}
I_{adj}\;\delta^{ab} = \sum_{a,b} f^{abc} f^{abd}
= h^\vee\psi^2 \delta^{ab}\,,
\end{equation}
where $\psi^2$ is the length squared of the longest root. We
will also need the trace of triple products of generators,
\begin{eqnarray}
{\rm Tr}_R \; (t^a t^b t^c) & = &
{1\over 2}\; {\rm Tr}_R([t^a,t^b] t^c)
+ {1\over 2}\; {\rm Tr}_R(\{t^a, t^b\} t^c)  \nonumber \\
& = & {i\over 2}  \, f^{abc}\, I_R
+ {1\over 2} \, d^{abc} \, I_R^{(3)}\, ,
\end{eqnarray}
where $d^{abc}$ is the totally symmetric invariant third order
tensor, and $I_R^{(3)}$ is related to the value in the
representation $R$ of the associated third order Casimir operator,
\begin{equation}
C_3 = \sum_{abc} d^{abc}\, t^a \, t^b\, t^c\ .
\end{equation}
Note that the $d^{abc}$ may vanish, as happens for example for
$su(2)$. The first non-trivial case, $su(3)$, will be  described
in more detail at the end of section~5.

  After normal ordering the expressions (\ref{R2}-- \ref{R4}),
{\it i.e.}\ moving all positive modes to the right of negative modes, 
and with the help of the above trace formulae we find~:
\begin{equation} \label{Rn2}
{\cal O}_{\rm reg}^{(2)}(R) = 2 \pi^2 I_R\;  J_0^a J_0^a \,,
\end{equation}
\begin{eqnarray}\label{Rn3}
{\cal O}_{\rm reg}^{(3)}(R)  &=&
{2\pi^3\over 3} \, I_R^{(3)}\,   d^{abc} \, J^a_0 J^b_0 J^c_0
\; + \;  4 \pi^2  I_R\; f^{abc} \sum_{n>0}
{1\over n}\;
J^a_{-n}J^b_0 J^c_n  \;  -
 \\
&&  \hspace*{-0.3cm}  -\;  4 \pi^2 i\; I_R\,  h^\vee\, \psi^2
\left[ \; 
\sum_{n>0} {1\over n} J^a_{-n} J^a_n \, -\, 
{1\over 2}\,  J^a_0 J^a_0  \left(
\sum_{n>0}{e^{-ns}\over n}  \right) \, + \,    
{\kappa\, \over 6}\, {\rm dim}(g) \; \left(\sum_{n>0} e^{-ns}
\right)  \right] 
\,, \nonumber
\end{eqnarray}
 and
\begin{eqnarray}\label{Rn4}
{\cal O}_{\rm reg}^{(4)}(R)   &=& : {\cal O}_{\rm reg}^{(4)}(R) :
-\;  2 \pi^2 \, I_R\,  h^\vee \, \psi^2\,\kappa\, \Biggl[\;
\sum_{n>0} {1\over n}\; J^a_{-n} J^a_{n}
  - \, J^a_0
J^a_0 \; \left( \sum_{n>0} {e^{-ns}\over n} \right) \, +
\nonumber \\
& & + \; { \kappa\over 4}\,   {\rm dim}(g)\,
\left(\sum_{n>0}e^{-ns} \right)
  \Biggr]\;\  + \;\ 
{\rm subleading}\,.
\end{eqnarray}
For the quartic operator we have not written out explicitly the
`subleading terms' that arise in the process of normal ordering if
one uses at least once the $if^{abc}$ piece  of the current
commutators. These terms have fewer powers of $J\sim \kappa^{1/2}$,
and they depend on the precise definition of  the
normal ordered expression  $:{\cal O}_{\rm reg}^{(4)}(R):$.
Indeed, different definitions of $:{\cal O}_{\rm reg}^{(4)}(R):$
differ by rearrangements of the positive, or of the negative
modes, which only involve the $if^{abc}$ piece of the 
commutator.

We have only calculated those `subleading' terms that  make divergent
contributions (in the $s\to 0$ limit) to the quadratic and cubic 
Casimir operators of $G$.\footnote{The ambiguities in the
normal-ordering prescription affect only the  finite subleading
terms.} Put differently, if we write  
\begin{equation}
{\cal O}_{\rm chir}^{\rm reg}(\lambda; R)\, = 
\, \cdots +  A_2(\lambda; R) \,J_0^aJ_0^a\,
+ \, A_3(\lambda; R) \,d^{abc}J_0^aJ_0^bJ_0^c +  \cdots  
  \ ,
\end{equation}
then the coefficients $A_2$ and $A_3$ read~: 
\begin{equation}\label{a2}
A_2(\lambda; R)\, = -2\pi^2 I_R \, 
\left[ \lambda^2 +  \lambda^3\xi +
\lambda^4 \left(\kappa\,\xi +  {3\over 4}\, \xi^2 + {\rm finite}\right)
+ O(\lambda^5)\right] \ 
\end{equation}
and 
\begin{equation}\label{a3}
A_3(\lambda; R)\, = \, -{2\pi^3 i\over 3}\; I_R^{(3)}\; \left[
\lambda^3 + \lambda^4\, \left({3\over 2}\, \xi + {\rm finite}\right)
+ O(\lambda^5)\right]\ , 
\end{equation}
where we have  defined
\begin{equation}\label{xi}
\xi = h^\vee\psi^2 \,{\rm log}\, s\, . 
\end{equation}
In calculating these results,  we made use of the infinite sums
\begin{equation}
\sum_{n=1}^\infty e^{- ns} = {1\over s } - {1\over 2} +
O(s)\ \ \ \quad \ \ \ \sum_{n=1}^{\infty} 
{e^{- n s}\over n} = -{\rm log}s + O(s) \ ,
\end{equation}
and dropped  terms that vanish as $s\to 0$. 

    Our discussion of regularised operators  can be extended easily to
the non-chiral defects of section 3. Regularising the operator 
${\cal O}_\omega(\lambda,\bar\lambda ; R)$ like 
${\cal O}_{\rm chir}(\lambda ; R)$ leads to the same expressions as 
(\ref{R2}--\ref{R4}), except for  the replacement (for say $\omega=1$)
\begin{equation}
 \lambda\tilde J_n^a \ \longrightarrow 
(\lambda J_n^a -  \bar\lambda\bar J_{-n}^a)\, 
e^{- |n|  s/2}\ .
\end{equation}
Left and right currents are  normal-ordered separately, giving
expressions like (\ref{Rn2}--\ref{Rn4}),  which we will not
explicitly write down. Note that the sum of frequencies in each term
does not have to add up to zero now, since non-chiral operators do   
not commute with the  Hamiltonian in the closed-string  channel.

\setcounter{equation}{0}
\section{Renormalisation and fixed points}

  Based on symmetry arguments, we expect that the divergent
contributions to the chiral operators of the previous section can be
absorbed into a redefinition of two parameters~: the coupling
$\lambda$  and an overall multiplicative factor.  These correspond to
the two local counterterms, proportional to the identity and to 
$J^a t^a$, that have dimension $\leq 1$ and respect the chiral and 
the global $G$-symmetries of the problem. Alternatively, the
only two background fields consistent with these
symmetries are a constant tachyon, and $A=Y$ proportional to the 
right-invariant one-form on $G$.  This huge reduction of parameter
space implies that by choosing $\lambda_{\rm eff}$ and $Z_R$ 
appropriately, we should be able to take the limit
\begin{equation}\label{rml}
{\cal O}_{\rm chir}^{\rm ren}(\lambda_{\rm eff}; R)
 \;    =\; {\rm lim}_{s\to 0}\; Z_R\; 
{\cal O}_{\rm chir}^{\rm reg}(\lambda; R)\  . 
\end{equation}
Close inspection of eqs.\ (\ref{path}) and (\ref{Rn2} -- \ref{a3})
shows that this is the case up to the order worked out in section~4.
A possible choice of the multiplicative renormalisation and of the
effective coupling, that removes all divergencies to this order, is~:  
\begin{equation}\label{eff}
\lambda_{\rm eff} = \lambda +
{1\over 2}(\lambda^2 + \kappa\lambda^3)\, \xi 
+ {1\over 4} \, \lambda^3\, \xi^2 + O(\lambda^4) \ ,
\end{equation}
and 
\begin{equation}\label{subtr}
{\rm log}\, Z_R = 2\,\pi^2\, C(R)\, h^\vee\,\psi^2 \,
\left(\, {1\over 3}\,\kappa\lambda^3 +  {1\over 4}\,\kappa^2\lambda^4
+ {\rm subleading} \right)\times {1\over s}\  , 
\end{equation}
where subleading stands for higher powers of $\lambda\sim
1/\kappa$ (this scaling is of interest for reasons that will be
discussed  in a minute). Note that, as anticipated in our notation, 
$\lambda_{\rm eff}$ is independent of the representation $R$, while
${\rm log} Z_R$ depends on it (to this order) via the eigenvalue of
the quadratic Casimir operator. Note also that $s$ is the ratio of the
only two length scales in the problem: the short-distance cutoff and
the circumference,  ${ L}$,  of the cylinder. Subtracting minimally
the $s$ pole, as in (\ref{subtr}),  is tantamount to a renormalisation
of the energy of the defect that only depends on the cutoff scale, 
and not on $L$. This same subtraction would have been automatically
implemented by the $\zeta$-function regularisation 
\begin{equation}
\sum_{n>0} 1 = \zeta(0) = - {1\over 2} \, .
\end{equation}
{}From equations (\ref{eff}) and (\ref{xi}) we can now extract the  
$\beta$-function 
\begin{equation}
\beta(\lambda_{\rm eff}) =  -{d\lambda_{\rm eff}\over d\, {\rm log}s}
=  - {1\over 2}h^\vee\psi^2 \left( \lambda_{\rm eff}^2 
+ \kappa \lambda_{\rm eff}^3 + O(\lambda_{\rm eff}^4)\right)
\, .
\end{equation}
The chiral defect is asymptotically free, and has an infrared  
fixed point at the critical value 
\begin{equation}\label{crit}
\lambda^* = -{1\over \kappa} + O\left({1\over \kappa^2}\right)\ .
\end{equation}
This  agrees  with the semiclassical argument of section 3
in the $\kappa >\!\!> 1$ region. The fact that the fixed point is
close to the origin justifies the use of perturbation theory in his
limit. Note, incidentally, that in the renormalisation scheme we 
have used the two-loop $\beta$-function is proportional to $\kappa$.

It is straightforward to extend this calculation to the 
general non-chiral defect (\ref{basic}). To leading order
in the $\lambda\sim\bar\lambda\sim 1/\kappa$ expansion, the effective
couplings in this case read
\begin{eqnarray}
\lambda_{\rm eff} & = & 
\lambda + {1\over 2}\, \xi\, 
    \left(\lambda^2 + \kappa (\lambda^3 + \lambda\, \bar\lambda^2)
    \right) +\cdots\nonumber \\
\bar{\lambda}_{\rm eff} & = & 
\bar\lambda  + {1\over 2}\, \xi\, 
    \left(\bar\lambda^2 + \kappa (\bar\lambda^3 + \bar\lambda\,
      \lambda^2) \right) + \cdots \ . \nonumber
\end{eqnarray}
The corresponding RG flow is  described by  the flow diagram of 
Figure~2.  As is manifest from this diagram, there are three
non-trivial fixed points: the stable chiral and anti-chiral fixed
points at $(\lambda,\bar\lambda)= (-1/\kappa , 0)$ and 
$(0,-1/\kappa)$,  as well as an unstable fixed point at
$\lambda=\bar\lambda=-1/2\kappa$.\footnote{The unstable fixed point
  corresponds probably to  D-branes  of the tensor product that
  decompose into D-branes of 
$\hat g_k\times \hat g_k/\hat g_{2k}$ and 
  $ \hat g_{2k}$, see ref. \cite{Quella:2002ct}. We thank Stephan Fredenhagen for a
 discussion of this point.} 
 This is again  in nice agreement
with the expectations based on the quasiclassical analysis of 
section~3.

\begin{figure}[ht]
       \hbox{\epsfxsize=90mm%
       \hfill~\hskip 3.5cm 
       \epsfbox{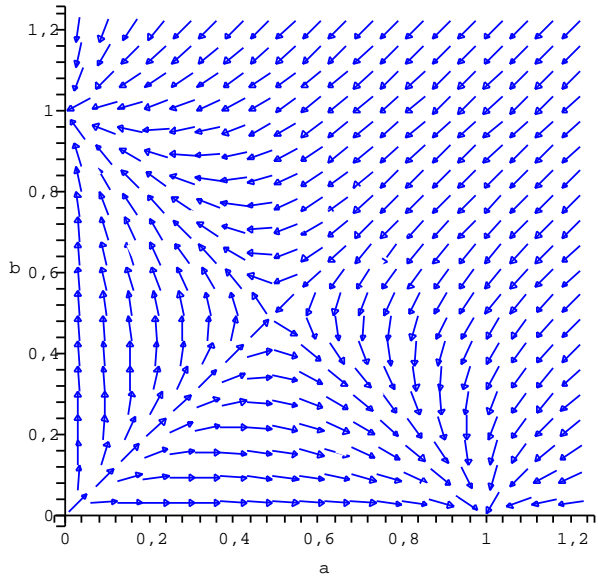} \hfill~}
       \caption{{\small The RG flow diagram}  
{\small for the general symmetric defects discussed in the
text. The space  of couplings is parametrised by} 
$(a, b) =-\kappa (\lambda , \bar\lambda)$. The stable fixed points
correspond to chiral defects, but there is also an unstable fixed
point along the line $\lambda =\bar\lambda$ .}
\end{figure}

\vskip 0.3cm

  As was furthermore argued in  section 3, the fixed-point operators
${\cal O}_{\rm chir}(\lambda^*; R)$ are non-trivial central
elements of the (left-moving) current algebra that take the value 
$S_{R\mu}/S_{0\mu}$ on the highest weight representation 
${\cal H}_\mu$. We now want to confirm these statements to the order
to which we have calculated 
${\cal O}_{\rm chir}^{\rm ren}(\lambda^* ;R)$, and this will occupy us  
in the remainder of the present section. The reader may choose to skip
these cumbersome calculations, and proceed directly to the following
section.  

Let us first verify to lowest order that 
${\cal O}_{\rm chir}^{\rm ren}(\lambda^* ; R)$ is central, {\it i.e.}\ 
that it commutes with all Kac-Moody generators $J_n^e$. Since at the
fixed point 
$\lambda = \lambda_{\rm eff} + O(\kappa^{-3})$, and since $Z_R$ is
only a multiplicative factor, it will be sufficient to check this for
the regularised bare expressions.  The contribution from the
second-order term is~:  
\begin{eqnarray} \label{o2com}
- {\lambda^2} [\, J^e_n\, , {\cal O}_{\rm reg}^{(2)}(R)]
& = & - 2  \pi^2 \lambda^2 \, I_R\, i  f^{eab}
\left( J^b_n J^a_0 + J^a_0 J^b_n \right) \nonumber \\
& = & - 4  \pi^2 \lambda^2\, I_R\, \left( i  f^{eab}
J^a_0 J^b_n + {1\over 2} \,  h^\vee \psi^2 \,  J^e_n\right) \, .
\end{eqnarray}
This needs to be cancelled by corresponding contributions from third
and fourth order. Since $\lambda \sim 1/\kappa$, the relevant
contributions  from the third-order term must be proportional to
$\kappa$. These must come from the central term of the commutator
involving $J^e_n$, and will only arise for $n\ne 0$ [for $n=0$ note
that (\ref{o2com}) vanishes, and so do all the other commutators
because of the global $G$-symmetry of the defect line]. 
The relevant contributions read
\begin{equation}\label{o3com}
- i {\lambda^3} [\, J^e_n\, ,{\cal O}_{\rm reg}^{(3)}(R)]
= - 4 \pi^2\, \lambda^3 \kappa\,  I_R
\left( i  f^{eab}\, J^a_0 J^b_n + h^\vee \psi^2 J^e_n \right)
+ \cdots  \,.
\end{equation}
Provided we choose $\lambda=-1/\kappa$, the term that is bilinear in
$J$ cancels. The term linear in $J$, on the other hand, is cancelled 
by a contribution at fourth order 
\begin{equation}\label{o4com}
{\lambda^4 } [\, J^e_n, {\cal O}_{\rm reg}^{(4)}(R)]
= - 2 \pi^2 \lambda^4 \, \kappa^2 \, I_R  \, h^\vee \psi^2 \, J^e_n
+ \cdots  \,.
\end{equation}
Thus, at the fixed point all  terms in (\ref{o2com}) -- (\ref{o4com}) 
cancel precisely, and there  are no further contributions  at order
$\kappa^{-2}$. Indeed, if it were to contribute at this order,  
the commutator of ${\cal O}_{\rm reg}^{(N)}(R)$ with $J^e_n$ would 
have to be proportional to $\kappa^{N-2}$. Since two $J$s are needed
to produce a power of $\kappa$, the only option is $N=5$, but then the 
resulting term would not involve any $J$s any more. This is
impossible because the total mode number of any  
${\cal O}_{\rm reg}^{(N)}(R)$ is  zero, so that the commutator with
$J^e_n$ must involve modes whose mode numbers sum to $n\ne 0$.

Let us now turn to the verification of the semiclassical formula
(\ref{topW}). Since 
${\cal O}_{\rm chir}^{\rm ren}(\lambda^* ; R)$ commutes with the
current algebra, it is sufficient to calculate its eigenvalue on  
the  (conformal) highest weight state $|\mu \rangle$ in
${\cal H}_\mu$. Using equations (\ref{Rn2}--\ref{Rn4}) and
(\ref{rml}--\ref{subtr}) we find~: 
\begin{eqnarray}
{\cal O}_{\rm
  chir}^{\rm ren}(\lambda ; R) \, |\mu\rangle & = &
\Biggl[\,  \dim(R)\, 
-\,  2  \pi^2 \, \lambda^2 \, I_R\, C(\mu)
\,  - \,  {2 i  \pi^3\over 3} \, \lambda^3 \, I^{(3)}_R \, C_3 (\mu)\,
  + \nonumber \\
& & \hskip -2cm 
+\,  
\pi^2\, 
\left( {  \kappa \over 3}\,\lambda^3 \,  +
{ \kappa^2 \over 4}\,\lambda^4 \,\right)
 I_R\, h^\vee\, \psi^2\,
\dim(g)\, 
 + \, {\rm subleading}\, \Biggr]\, |\mu\rangle  
 \,, \label{evalu}
\end{eqnarray}
where we are neglecting  terms of order $\kappa \lambda^4 J$ and 
$\lambda^4 J^3$, and of course all terms of order $\lambda^5$.
Note that $C(\mu)$ and  $C_3(\mu)$ denote the values of the quadratic
and cubic Casimir operators  in the representation $\mu$ (see 
section 4). 

\noindent Plugging into this formula the critical value
(\ref{crit}) one  finds~: 
\begin{equation}\label{qdim}
{\cal O}_{\rm chir}^{\rm ren}(\lambda^* ; R)\, |\mu \rangle =
\left[  \dim(R)   - {2\pi^2 \over \kappa^2} \, I_R\, C(\mu)
- {\pi^2 \over 12 \kappa^2} \, I_R\, h^\vee\, \psi^2\,
\dim(g) + { O}\left({1\over \kappa^3}\right) \right]\, |\mu \rangle \,.
\end{equation}
The expression in brackets  agrees  precisely with
the expansion of the generalised quantum dimension
${S_{R\mu}/ S_{0\mu}}$ up to cubic order.\footnote{We thank 
Daniel Roggenkamp and Terry Gannon for helping us check this in
general.}
We can in fact do still a little better: 
the terms that we did not calculate
explicitly in (\ref{evalu}) cannot contribute to the coefficient of
the quadratic or cubic Casimir at order $\kappa^{-3}$ (but can only
modify the constant term at this order). If we assume that the
critical value of the coupling constant is shifted in the familiar way,
\be\label{critc}
\lambda^\ast = - {1\over \kappa + h^\vee\psi^2/2 } +
O\left({1\over \kappa^3} \right)\ , 
\ee
then (\ref{evalu}) actually reproduces the coefficients of the
quadratic and cubic Casimir of ${S_{R\mu}/ S_{0\mu}}$   even up to
cubic order, at least for $su(2)$ and $su(3)$.  For $su(2)$ this is
simply a consequence of the fact that $I_R^{(3)}=0$, and that the
generalised quantum dimensions are even functions of $k+2$. In the
case of $su(3)$, we use the fact  (see for example \cite{Bais})~: 
\begin{equation}
 d^{abc} {\rm Tr}_R(t^a t^b t^c) = \dim(R)\, C_3(R) =
I_R^{(3)} \,  d^{abc}\, d^{abc} =
{40\over 3} \, I_R ^{(3)}\, ,
\end{equation}
where we have here set $\psi^2=1$. 
It follows that the order $\kappa^{-3}$ contribution of
(\ref{evalu}) at the critical value of the coupling
(\ref{critc}) is equal to~:
\begin{equation}\label{dynk}
-  {i\, \pi^3 \over 20\, \kappa^3}\, \dim(R)\, C_3(R)\, C_3(\mu) \,.
\end{equation}
If we describe the representation $\mu$ by the Dynkin labels
$[\mu_1,\mu_2]$, then the cubic Casimir operator reads
(see for example reference  \cite{AM}, which uses however a different 
normalisation convention from ours)~:
\begin{equation}
C_3(\mu) = {1\over 36}\, (\mu_1-\mu_2)\,
(\mu_1+2\mu_2+3)\, (\mu_2+2\mu_1+3) \,.
\end{equation}
Substituting this into (\ref{dynk}) reproduces  the cubic contribution
of the generalised quantum dimension for $su(3)$, as can be checked
explicitly  by using the description of the $S$-matrix elements given
in ref.\ \cite{Gannon}.

\setcounter{footnote}{0}
\section{Boundary RG flows}

  Finally, we can now return to our original problem, which concerned
the renormalisation group  flow  of boundary states. The main point 
 we want  to stress  is the following~: since the renormalised loop
operators have been  constructed  in the bulk, their action on the
boundary defines  {\it universal} RG  flows, independent of the
ultraviolet brane  $n\vert {\cal  B}\gg$.\footnote{Assuming of course
that the action  is non-singular. As we  have discussed in the
previous sections, this is guaranteed for the  chiral loop operators
constructed here.} Conformal defect lines, in particular, define
(generally non-invertible) maps in the space of conformal boundary
states, {\it i.e.}\ they act  as solution-generating symmetries of
open string theory.  What this means 
geometrically is that there exist universal bundles,
$A$ and $Y$,  on the target space which solve the open-string
equations if pulled back on {\it  any} classically consistent
brane. Note that unlike the continuous or discrete automorphisms of
the CFT, these maps change in general the tension (or $g$-function)  
of the D-brane. Note also that the universality of the flows may shed
light on the empirical observation that the charge groups of the
untwisted and twisted D-branes always coincide \cite{GG2}.

  The endpoint of the boundary RG flow induced by the symmetric WZW
defects of this paper, is given by the action of the loop operator at
one of the three non-trivial fixed points of figure 2. Let us consider
first the chiral case.  Most of what 
we describe in the following is well known to the experts (see for 
example \cite{Petkova:2000ip,Graham:2003nc}), but we include it for 
completeness.  The space of states of the closed string theory describing
strings on the simply-connected group manifold $G$ is
\begin{equation}\label{spectrum}
{\cal H} = \bigoplus_{\mu\in P^+_k(g)} {\cal H}_{\mu}
\otimes \bar{\cal H}_{\mu^\ast}\,,
\end{equation}
where $P^+_k(g)$ denotes the set of integrable highest weight
representations of the affine algebra $\hat{g}$ at level $k$. The
D-branes that preserve the affine symmetry up to the (possibly
trivial) automorphism $\omega$  are characterised by the gluing
condition 
\begin{equation}\label{gluing}
\left[J^a_n + \omega\left(\bar{J}^a_{-n}\right) \right]\, |\!|
\alpha \rangle\!\rangle = 0 \,.
\end{equation}
As always, these D-branes can be expanded in terms of the
corresponding Ishibashi states
\begin{equation}
|\!|\alpha\rangle\!\rangle = \sum_{\mu\in{\cal E}_\omega}
{\psi_{\alpha \mu} \over \sqrt{S_{0\mu}}} \,
|\mu \rangle\!\rangle^\omega \,,
\end{equation}
where $|\mu\rangle\!\rangle^\omega$ is the unique Ishibashi state
satisfying (\ref{gluing}) in the sector
${\cal H}_{\mu} \otimes \bar{\cal H}_{\mu^\ast}$, and 
${\cal  E}_\omega$ denotes the set of exponents, {\it i.e.} 
\begin{equation}
{\cal E}_{\omega} = \left\{ \mu\in P^+_k(g): \omega(\mu) = \mu
\right\} \,.
\end{equation}
The D-branes are then uniquely characterised by the unitary matrix
$\psi_{\alpha \mu}$. For an introduction to these matters see for
example \cite{BPPZ}.

For the case $\omega={\rm id}$, ${\cal E}_\omega = P^+_k(g)$, and
the D-branes are naturally labelled by the integrable highest
weight representations of $\hat g$. The matrix
$\psi_{\alpha \mu}=S_{\alpha \mu}$ is then the modular $S$-matrix,
and the corresponding D-branes are sometimes referred to as the
`Cardy branes'. If $\omega$ is non-trivial, the D-branes are
often called the `twisted' D-branes; they are naturally
labelled by the $\omega$-twisted representations of the affine
algebra, and the $\psi$-matrix is then the modular transformation
matrix describing the modular transformation of twisted and twined
characters \cite{BFS,GG1}. In either case, the open
string that stretches between the branes labelled by $\beta$ and
$\alpha$ then contains the representation ${\cal H}_\nu$ of $\hat{g}$
with multiplicity
\begin{equation}\label{nim}
{\cal N}_{\nu \alpha}{}^{\beta} = \sum_{\mu\in {\cal E}_\omega}
{\psi_{\alpha \mu} \, S_{\nu \mu} \, \psi_{\beta \mu}^\ast \over
S_{0\mu}} \,.
\end{equation}
The consistency of the construction requires that these numbers
are non-negative integers, and in fact, they must define a
NIM-rep of the fusion algebra -- see {\it e.g.}\
\cite{BPPZ,gannon} for an introduction to these matters.
\medskip

We are interested in studying the action of the chiral 
loop operator on the boundary state labelled by $\alpha$. Since the
loop operator only involves the left-moving modes $J^a_n$, it
obviously commutes with the right-moving modes $\bar{J}^a_m$ for any
value of the (renormalised) coupling constant $\lambda$. We have
argued above that for the critical coupling, $\lambda=\lambda^\ast$,  
the operator also commutes with the left-moving currents $J^a_n$. Thus
at the critical coupling, the loop operator maps a boundary state
satisfying (\ref{gluing}) to another boundary state satisfying
(\ref{gluing}). We want to calculate this resulting boundary
state. 

At the critical coupling, the loop operator commutes with all modes
$J^a_n$, and therefore must act as a $C$-number on each irreducible
representation ${\cal H}_\mu$ of $\hat{g}$. As we have argued
before, this $C$-number is precisely equal to the generalised quantum
dimension  
\begin{equation}\label{cnum}
\left. {\cal O}_{\rm chir}^{\rm ren}(\lambda^* ; R)
\right|_{{\cal H}_\mu} =
{S_{R\mu}\over S_{0\mu}} \,.
\end{equation}
Thus we find
\begin{equation}\label{loopa}
{\cal O}_{\rm chir}^{\rm ren}(\lambda^* ; R) \,
|\!|\alpha\rangle\!\rangle = \sum_{\mu\in{\cal E}_\omega}
{S_{R \mu} \over S_{0 \mu}} \, {\psi_{\alpha \mu}\over
\sqrt{S_{0 \mu}}} \, |\mu\rangle\!\rangle^\omega \,.
\end{equation}
For each $\mu\in{\cal E}_\omega$ we write
\begin{eqnarray}
{S_{R \mu} \over S_{0 \mu}} \, {\psi_{\alpha \mu}\over
\sqrt{S_{0 \mu}}} & = & \sum_{\beta} {\psi_{\beta \mu} \over
\sqrt{S_{0 \mu}}} \sum_{\nu\in {\cal E}_\omega}
{\psi^\ast_{\beta \nu} S_{R \nu} \psi_{\alpha \nu} \over S_{0 \nu}}
\nonumber \\
& = & \sum_{\beta} {\psi_{\beta \mu} \over \sqrt{S_{0 \mu}}} \,
{\cal N}_{R \alpha}{}^{\beta} \,,
\end{eqnarray}
where the unitarity of $\psi_{\alpha \nu}$ was used in the first
line, {\it i.e.}
\begin{equation}
\sum_{\beta} \psi^\ast_{\beta \nu} \, \psi_{\beta \mu} =
\delta_{\nu\mu} \,,
\end{equation}
with both $\mu$ and $\nu$ elements of ${\cal E}_\omega$. In the second 
line we have inserted the definition of the NIM-rep coefficient
(\ref{nim}).
Putting this back into (\ref{loopa}) we then find
\begin{equation}
{\cal O}_{\rm chir}^{\rm ren}(\lambda^* ; R) \,
|\!|\alpha\rangle\!\rangle = \sum_{\beta} 
{\cal N}_{R\alpha}{}^{\beta} \, |\!|\beta\rangle\!\rangle \,. 
\end{equation}
This therefore reproduces  the flow that was proposed in
\cite{Fredenhagen:2000ei}, based on the
analysis of the non-commutative worldvolume
actions \cite{Alekseev:2000fd,Alekseev:2002rj} and the Kondo problem 
\cite{AL,Affleck:1995ge}. These flows were used in
\cite{Fredenhagen:2000ei,Maldacena:2001xj,Bou} to show that the
untwisted branes carry the charge that was predicted by K-theory; for
the twisted D-branes this was shown in \cite{GG2}. 
\smallskip

  In the diagonal theories discussed above, the actions  of the
chiral and antichiral operators coincide. This is also true
(semiclassically at least) for the non-chiral operator  
${\cal O}_\omega^{\rm ren}(\lambda^\ast/2 ,\lambda^\ast/2 ; R) $, if  
the  gluing automorphism of the ultraviolet brane is the same as 
the one that goes  into the definition of the defect line.  
More generally, the combination of defect and boundary breaks  the
global $G$ symmetry of the problem.  The pull-back of the defect 
to the boundary may, in this case, be singular, corresponding to the
excitation of modes that take us outside the original 
two-parameter space of couplings. 
This raises the  more general
problem   of the fusion between arbitrary defect and boundary flows;
for topological defect lines  this question  was analysed
in  ref.\ \cite{Graham:2003nc}.

The analysis of this paper   can be extended easily to the D-branes of
any bulk theory in which left and/or right current algebras exist. 
Examples include   WZW models of non-simply connected
group manifolds,\footnote{A systematic analysis of D-brane charges for
these theories was begun in \cite{GG3}, see also \cite{BS,Fred}.}
as well as  WZW  orbifolds and coset models in which part of the
current algebras survives. It is less clear
whether the above ideas can be applied to more general  coset models, 
or to the D-branes of  WZW models for  non-compact groups. These
questions deserve further investigation.

\vfil\break 

\centerline{\large \bf Acknowledgements}
\vskip .2cm

This research has been  partially supported by  the European Networks 
`Superstring Theory' (HPRN-CT-2000-00122) and  `The Quantum Structure
of Spacetime' (HPRN-CT-2000-00131), as well as by the Swiss National
Science Foundation. We have benefited from discussions with 
Anton Alekseev, Giovanni Felder, Stefan Fredenhagen, J{\"u}rg
Fr{\"o}hlich, Terry Gannon, Kryzstof Gawedzki, Hanno Klemm, Ruben
Minasian, Boris Pioline, Ingo Runkel, Volker Schomerus, Samson
Shatashvili, Jan Troost and Sasha Zamolodchikov. We also thank the
organisers of the Ascona workshop, where part of the work has been
carried out.

\end{document}